\documentclass[twocolumn,showpacs,a4paper,superscriptaddress,affilletter,amsmath,amssymb]{revtex4}

\usepackage{layouts}
\usepackage{graphicx}

\begin{document}


\title{p-type doping of II-VI heterostructures from surface states:
 application to ferromagnetic Cd$_{1-x}$Mn$_x$Te quantum wells}%

\author{W. Ma\'slana}
\affiliation{Institute of Experimental Physics, Warsaw University,
Ho\.za 69, 00-681 Warsaw, Poland} \affiliation{Laboratoire de
Spectrom\'etrie Physique, CNRS et Universit\'e Joseph
Fourier-Grenoble, B.P.87, 38402 Saint Martin d'H\`{e}res Cedex,
France} 

\author{M. Bertolini}
\author{H. Boukari}
\affiliation{Laboratoire de Spectrom\'etrie Physique, CNRS et
Universit\'e Joseph Fourier-Grenoble, B.P.87, 38402 Saint Martin
d'H\`{e}res Cedex, France}

\author{P. Kossacki}

\affiliation{Institute of Experimental Physics, Warsaw University,
Ho\.za 69, 00-681 Warsaw, Poland}

\affiliation{Laboratoire de Spectrom\'etrie Physique, CNRS et
Universit\'e Joseph Fourier-Grenoble, B.P.87, 38402 Saint Martin
d'H\`{e}res Cedex, France}

\author{D. Ferrand}
\affiliation{Laboratoire de Spectrom\'etrie Physique, CNRS et
Universit\'e Joseph Fourier-Grenoble, B.P.87, 38402 Saint Martin
d'H\`{e}res Cedex, France}

\author{J. A. Gaj}

\affiliation{Institute of Experimental Physics, Warsaw University,
Ho\.za 69, 00-681 Warsaw, Poland}

 \author{S. Tatarenko}
 \author{J. Cibert}
\affiliation{Laboratoire de Spectrom\'etrie Physique, CNRS et
Universit\'e Joseph Fourier-Grenoble, B.P.87, 38402 Saint Martin
d'H\`{e}res Cedex, France}

\date{\today}

\begin{abstract}
We present a study of p-type doping of CdTe and Cd$_{1-x}$Mn$_x$Te
quantum wells from surface states. We show that this method is as
efficient as usual  modulation doping with nitrogen acceptors, and
leads to hole densities exceeding $2 \times 10^{11}$ cm$^{-2}$.
Surface doping was applied to obtain samples with
Cd$_{1-x}$Mn$_x$Te quantum well with  up to $x=9.3$\% containing
hole gas. We could also increase the growth temperature up to
280$^\circ$C, which results in sharper photoluminescence lines,
when compared to the similar nitrogen doped samples.
Carrier-induced ferromagnetism was observed in surface doped
samples.

\end{abstract}



\maketitle

 Studies of carrier induced ferromagnetism in
diluted magnetic semiconductors (DMS) are important for the
development of spintronics - a new field exploiting spin degrees
of freedom for information processing. The most conclusive results
have been obtained for p-type doped structures. In III-V DMS, the
same impurity (Mn) carries the localized spins and acts as an
acceptor~\cite{Ohno98} which puts strong limits on the realization
of QWs and 2D systems~\cite{Ohno00} using thin layers of III-V
DMS~\cite{Boseli00,Byo00}. In Cd$_{1-x}$Mn$_x$Te quantum wells
(QW), it has been proposed theoretically \cite{Dietl97} and shown
experimentally\cite{Haur97} that the presence of a two dimensional
(2D) carrier gas can induce a ferromagnetic ordering due to the
strong exchange coupling of Mn spins to charge carriers. In
\cite{Haur97}, the modulation-doped structures were grown by
molecular beam epitaxy using nitrogen acceptors in the
Cd$_{1-y-z}$Zn$_{z}$Mg$_{y}$Te barriers. Although efficient, this
method brings certain restrictions. Effective doping of
Cd$_{1-y-z}$Zn$_{z}$Mg$_{y}$Te with nitrogen is only possible for
$y$ lower than 30\%, it requires lowering the growth temperature
from the usual 280$^\circ$C down to 220$^\circ$C to avoid a strong
interdiffusion of the heterostructure \cite{Arn98}, and thus
affects the quality of the samples, e.g., by increasing interface
roughness~\cite{Grie94}. Moreover the presence of nitrogen
precludes almost any post-growth treatment of the sample.

In this letter we present a method of p-type doping of
Cd$_{1-x}$Mn$_x$Te QWs from surface states. We show that this
method can be as efficient as modulation doping with nitrogen in
supplying a hole gas to the QWs, and can be used to induce a
ferromagnetic order in the QW. Furthermore it increases the
thermal stability of structures  allowing growth and processing of
samples at higher temperature. Finally, it makes possible to
obtain a hole gas in deeper quantum wells or quantum wells with a
higher Mn content.

Samples have been grown by molecular-beam epitaxy on two types of
(001) substrates, Cd$_{0.88}$Zn$_{0.12}$Te which is transparent at
the energy of the QW transition, and Cd$_{0.96}$Zn$_{0.04}$Te. For
comparison purposes the growth parameters were first kept in the
range typical for the growth of the nitrogen doped samples,
including the substrate temperature 220$^\circ$C. Each sample
contained a 100\AA~wide QW made of Cd$_{1-x}$Mn$_{x}$Te or CdTe.
The QW was embedded between Cd$_{0.7}$Zn$_{0.08}$Mg$_{0.22}$Te
barriers in case of 12\% Zn substrate and Cd$_{0.78}$Mg$_{0.22}$Te
for the samples grown on the 4\% Zn substrate, so that the whole
structure could be grown coherently strained to the substrate. The
thickness of the top barrier was in the range from 150\AA~to
1000\AA. The barrier on the substrate side was 3000\AA~ thick.

\begin{figure}[b!]
  \begin{center}
\includegraphics[width=.45\textwidth]{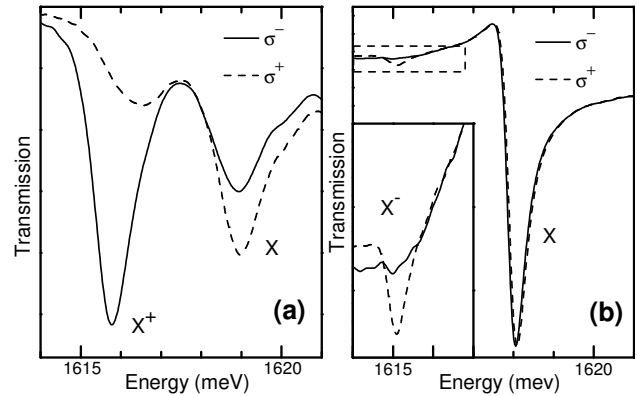}
  \end{center}
\caption{ Circularly polarized transmission spectra, at 3T and
1.7K, for a single CdTe QW with a cap layer thickness of (a)
250\AA~and (b) 1000\AA. The inset in (b) presents a close-up of
the area of the spectrum attributed to a negatively charged
exciton. Note the change in lineshape which is consistent with the
change in the cap layer thickness} \label{Fig1}
\end{figure}

All properties discussed below were determined by magneto-optical
spectroscopy in the Faraday configuration  (magnetic field
perpendicular to the sample surface), with the sample mounted
strain-free in liquid helium in a superconducting magnet. The
experimental setup allowed us to perform transmission and
reflectivity studies using a halogen lamp, and photoluminescence
(PL) and PL excitation (PLE) using a tunable Al$_2$O$_3$:Ti laser
providing about 2 mW/cm$^2$. The composition of barriers was
checked from the PL transition energy \cite{Hart96}. The Mn
content in the QW was checked from fitting a modified Brillouin
function \cite{Gaj94} to the Zeeman splitting measured in PL.

\begin{figure*}[bt]
\begin{center}
\includegraphics[width=.8\textwidth]{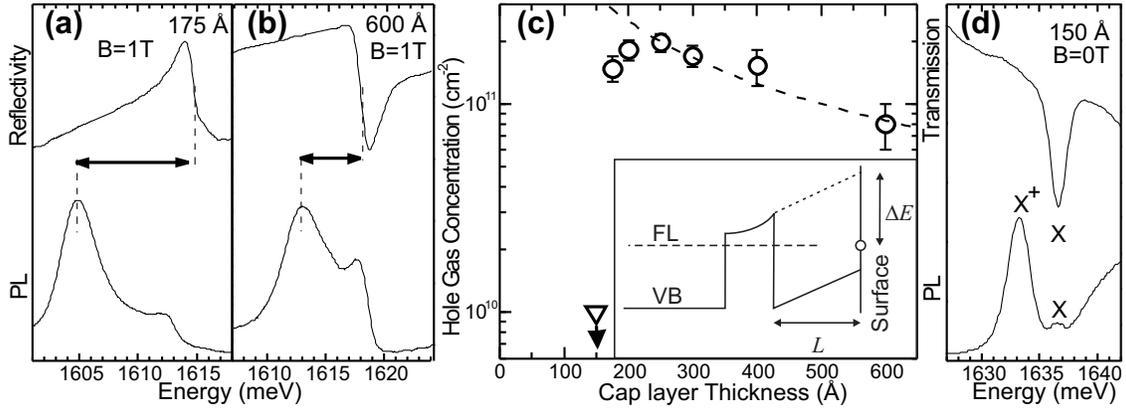}
  \end{center}
 \caption{
\textbf{(a),(b)} Reflectivity (top)  an PL spectra (bottom) at 1.7
K and 1 T for two Cd$_{0.99}$Mn$_{0.01}$Te QWs with different cap
layer thicknesses, as indicated. Arrows note the Moss-Burstein
shift.
 \textbf{(c)} Hole density in 100\AA~wide
Cd$_{0.99}$Mn$_{0.01}$Te QWs, determined from the Moss-Burstein
shift at 1T (circles) or from the charged exciton intensity
(triangle, two samples). The dashed line is the density expected
from surface acceptor states at energy $\Delta E=90$meV from the
QW state (model schematized in the inset: FL - Fermi level, VB -
Valence band, $L$- distance of the acceptor sheet from the QW.
 \textbf{(d)} Transmission (top) and PL spectra (bottom)
for a Cd$_{0.99}$Mn$_{0.01}$Te QW with a 150\AA~thick cap layer.
Neutral and charged exciton lines dominate the spectra indicating
a low carrier density. }\label{capfig}
\end{figure*}

The first evidence for the presence of a  carrier gas in a QW
close to the surface comes from transmission spectra with applied
magnetic field (Fig.~\ref{Fig1}a). Two narrow absorption lines are
observed. Their splitting (about 3 meV) and the strong circular
polarization of the low energy line are a fingerprint of a charged
exciton transition, and therefore an indication of the presence of
a carrier gas \cite{Khen93, Haur96, Koss99}.

In a CdTe QW subject to a compressive  biaxial strain, such as
those here which were coherently grown on a
Cd$_{0.88}$Zn$_{0.12}$Te substrate, the sign of the charge
carriers is unambiguously determined from the circular
polarization of the charged exciton absorption in Faraday
configuration: Due to the same signs of the electron and hole
g-factors, the positively charged exciton is observed in
$\sigma^-$ polarization\cite{ Haur96} while the negatively charged
exciton is observed in $\sigma^+$ polarization \cite{Khen93}. In
Fig.~\ref{Fig1}a, the lower energy line obeys the selection rules
for positively charged excitons, which shows that free holes are
present in this CdTe QW close to the surface. Note that no
nitrogen acceptors were introduced into this sample. The opposite
polarization rule was observed in a CdTe QW identical to the
previous one but with a thicker (1000\AA) cap layer
(Fig.~\ref{Fig1}b). The weak intensity of the X$^-$ line indicates
a low concentration of electrons. Hence we conclude that we have a
weak residual doping, which is n-type and accounts for the low
electron density in the deeply buried QW, while the hole gas with
a higher density in the QW close to the surface has a different
origin.

\begin{figure}[b!]
  \begin{center}
 \includegraphics[width=.45\textwidth]{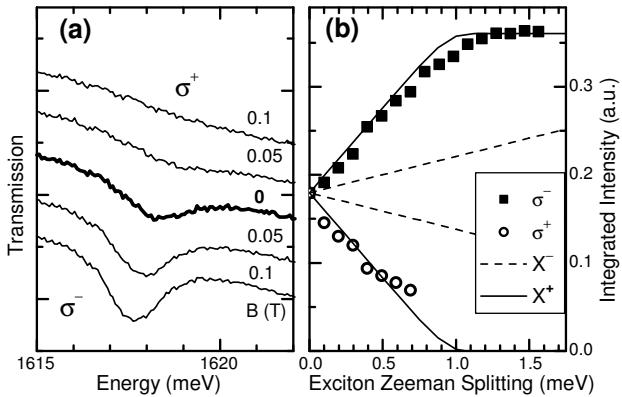}
  \end{center}
\caption{\textbf{(a)} Transmission spectra, in magnetic field, for
a Cd$_{0.995}$Mn$_{0.005}$Te QW placed 250\AA~below the surface of
the sample. Measured hole
concentration is $2\times 10^{-11}$cm$^{-2}$. 
\textbf{(b)} Integrated intensity of transmission line versus
 Zeeman splitting in both circular polarizations (points)
and results of calculation for electron (solid line) and hole
gases (dashed line).} \label{Fig3}
\end{figure}

The presence of a carrier gas was also  evidenced in
Cd$_{1-x}$Mn$_x$Te QWs. In samples with a thin cap layer, the
so-called Moss-Burstein shift between the absorption (or
reflectivity or PLE) line and the PL line indicates a large
density of carriers in the QW (Fig.~\ref{capfig}a, b). In the case
of band-to-band transitions, the Moss-Burstein shift is equal to
the sum of the kinetic energies of electrons and holes (which are
involved in the excitation process) at Fermi wavevector $k_F$, so
that it can be used as a tool for measuring the carrier density.
For a better accuracy, the Moss-Burstein shift was measured at a
magnetic field such that the carrier gas was fully polarized: Then
the Moss-Burstein shift measured in $\sigma^+$ polarization is
twice its value at zero field. Note the change in reflectivity
line in agreement with the cap layer thickness \cite{Zheng88}. The
carrier density was then calculated assuming an effective electron
mass $m^*_e=0.1 m_0$, and an in-plane hole mass $m^*_h=0.25 m_0$
\cite{Fish95}. We estimate that the relative carrier concentration
is then determined to within a few \%, while its absolute value is
accurate to within a factor of two due to residual excitonic
effects \cite{Koss99}. The determination of the carrier density
from the Moss-Burstein shift makes no assumption on the nature
(electrons or holes) of the carriers. However, once the density of
carriers is known, their sign can be deduced from the value of the
magnetic field necessary to fully polarize the carrier gas. Full
polarization is witnessed by the vanishing of the charged exciton
absorption in $\sigma^+$ polarization(Fig.~\ref{Fig3}). Due to the
giant Zeeman effect, characteristic for diluted magnetic
semiconductors (DMS), complete spin polarization of a hole gas of
density $2\times 10^{-11} $cm$^{-2}$ in a Cd$_{0.99}$Mn$_{0.01}$Te
QW is achieved when applying a magnetic field as low as 0.1T. This
field is expected to be at least 5 times larger for the same
density of electrons\cite{Gaj94,Koss99}.

A set of samples with similar growth conditions,  and almost
identical structure, allowed us to investigate the influence of
the cap layer thickness on the QW hole density. In this series of
samples, contrary to samples shown in Fig.~\ref{Fig1}, we used a
thin layer of nitrogen doped barrier material, 1000\AA~below the
QW, to screen any spurious effect from the interface with the
substrate. The hole density (as deduced from the Moss-Burstein
shift) significantly increases when the thickness of the cap layer
decreases from 600 to 250\AA~(Fig.~\ref{capfig}c). This reinforces
the idea that the origin of the hole gas is not linked to a
residual doping of the material, but is due to the presence of
electron traps (acceptors) on the surface. We can calculate the
hole density expected in the QW assuming a high density of
acceptor states on the surface, with the energy position of these
surface states as the only adjustable parameter. If we define this
position by the distance $\Delta E$ between the acceptor state and
the level confined in the QW, the hole density is mostly
determined by the effect of the electric field between the QW and
the surface (see the inset in Fig.~\ref{capfig}c), \textit{i.e.}
we can neglect small contributions such as the change of
confinement energy within the QW, the kinetic energy of confined
carriers, or the valence band shift due to the small amount of Mn
in the QW. Then the QW hole density simply writes $
  p=\varepsilon\varepsilon_0 \Delta E / (e L)
$
where $\varepsilon$ (=10) is the  dielectric constant of the cap
material, $\varepsilon_0$ is the permittivity of vacuum, $e$ is
the electron charge, and $L$ is the thickness of the cap layer. A
good fit is obtained if we take $\Delta E$ = 90 meV (dashed curve
in Fig.~\ref{capfig}c).

This value should be considered with care. First  the carrier
density is probably higher than the value we determine optically,
since illumination tends to decrease the carrier density, even
when using a low intensity at photon energy smaller than the
barrier gap. In addition, one should keep in mind that - due to
residual excitonic effects - the Moss-Burstein shift determination
is not too precise on the absolute value of the carrier
density\cite{Koss99}. Also, a  drop of the carrier density is
observed on the samples with a very thin cap layer (below 200\AA).
The decrease is dramatic in the two samples grown with $L=
150$\AA, where the free exciton is observed in transmission
(Fig.~\ref{capfig}d), so that we estimate the carrier density to
be smaller than $10^{-10}$cm$^{-2}$ \cite{Koss00jc}. Further
studies are needed to elucidate the origin of this sharp drop.

\begin{figure}[bt]
   \begin{center}
 \includegraphics[width=.45\textwidth]{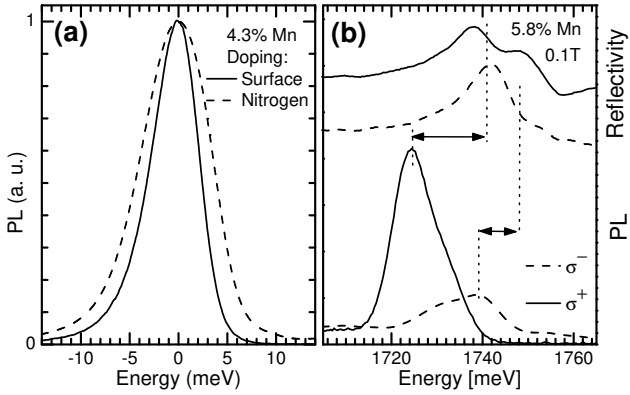}
  \end{center}
 \caption{
 \textbf{(a)} Zero field PL spectra of a single
Cd$_{0.957}$Mn$_{0.043}$Te QW at 4.2K. The QW is depleted from
carriers using blue light illumination. One sample (dashed line)
was grown at  220$^\circ$C and doped with nitrogen. The other one
was doped from the surface states (250\AA~cap layer) and  grown at
280$^\circ$C. Both have the same Mn content and carrier density in
the dark.  The spectra are normalized and their positions are
shifted.
 \textbf{(b)} PL and reflectivity spectra  in circular polarizations
at 0.1T and 1.7K  for a sample with 5.8\%Mn in the QW.
}\label{Fig4}
\end{figure}

The main result is that  we obtained hole densities exceeding
$2\times 10^{11}$cm$^{-2}$ in the QWs with a cap layer thickness
of 250\AA~without using impurities.  The presence of holes
probably also explains PL spectra observed in nominally undoped
parts of gradually doped samples~\cite{Wojt00}.
 We show now that this allows
us to increase the growth temperature and to grow DMS QWs with a
higher Mn content than our previous upper limit (below $x=0.05$).

Undoped samples have been shown to exhibit sharper lines when
grown at higher temperatures \cite{Grie94} due to reduction of
interface roughness. In doped samples other cases of broadening
can operate. However in figure ~\ref{Fig4}a we compare PL spectra
of a nitrogen doped sample grown at 220$^\circ$C, and of a sample
doped from the surface states and grown at 280$^\circ$C. Both
samples contain 4.3\% Mn in the QW, and identical carrier
densities in the dark. We present spectra measured in zero
magnetic field and under blue illumination in order to deplete the
QW from hole gas. Sharper lines are observed in the second sample.

Carrier induced ferromagnetism was  observed in those samples with
4.3\% Mn \cite{26ICPS:Maslana}, with the same behavior in the
nitrogen and surface doped samples.
 In addition, we can grow Cd$_{1-x}$Mn$_x$Te QWs with the $x$ up
 to 9.3\% and yet a significant hole density. As an example
 figure \ref{Fig4}b presents the PL and reflectivity spectra in circular polarizations
for a sample with 5.8\% Mn in the QW at magnetic field of 0.1T, so
that the hole gas is fully polarized. The shift in $\sigma^-$
polarization can be attributed to disorder (Stokes shift). The
shift in $\sigma^+$ polarization is significantly higher due to
the presence of the hole gas and accompanying Moss-Burstein shift.

We have no conclusive information  on the nature of the surface
states involved in the formation of the hole gas. The presence of
acceptor states at the surface of Cd$_{1-x}$Zn$_{x}$Te has been
reported by Yang et al. \cite{Yang02} and attributed to the
formation of TeO$_2$. In order to control the formation of oxides
on the Cd$_{1-y-z}$Mg$_y$Zn$_z$Te surface, a 50\AA~ thick layer of
amorphous tellurium can be deposited at --20$^\circ$C right after
the growth. We used two samples placed side-by-side on the
substrate holder during the growth. Then one sample was removed
from the MBE chamber, and the second one was heated up to
240$^\circ$C for a few seconds in the vacuum, in order to
re-evaporate the Te layer (as controlled by RHEED). No hole gas
was present in the QW of the sample protected by the Te layer,
while a hole density $2\times10^{11}cm^{-2}$ was found in the QW
of the sample without the Te cap. This suggests that surface
oxides may indeed play a role in the formation of the electron
traps.

In conclusion, we  demonstrate an efficient method for doping CdTe
QWs using surface acceptor states. Hole densities in excess of
$2\times 10^{11}$ cm$^{-2}$ in the QW have been measured by
optical spectroscopy. The investigated mechanism was applied to
obtain samples containing up to 9.3\%  Mn in the QW with a
significant hole gas density. Efficient doping can be achieved
without the technological restrictions arising from incorporating
nitrogen impurities into the structures. We can therefore use
higher temperatures for the growth of the samples (and  further
processing) and obtain a hole gas in deeper quantum wells or
quantum wells with a higher Mn content.


This work is partially supported by the KBN grant 5P03B02320
 and Polish-French Collaboration Program Polonium.


\bibliography{Art_Doping}

\end{document}